\documentclass[sigconf]{acmart}
\AtBeginDocument{%
  \providecommand\BibTeX{{%
    \normalfont B\kern-0.5em{\scshape i\kern-0.25em b}\kern-0.8em\TeX}}}
    
\setcopyright{acmcopyright}
\copyrightyear{2020}
\acmYear{2020}
\acmDOI{10.1145/1122445.1122456}

\acmConference[EICC '20]{EICC '20: ACM Symposium on European Interdisciplinary Cybersecurity Conference}{November 18--19, 2020}{Rennes, France}
\acmBooktitle{EICC '20: ACM Symposium on European Interdisciplinary Cybersecurity Conference, November 18--19, 2020, Rennes, France}
\acmPrice{15.00}
\acmISBN{978-1-4503-XXXX-X/18/06}
\usepackage{amsmath,amssymb,amsthm}
\begin{document}

\title[Bayesian Trend Analysis of Cyberbullying]{Evaluating the Impact of COVID-19 on Cyberbullying through Bayesian Trend Analysis}

\author{Sayar Karmakar}
\affiliation{%
  \institution{University of Florida}
    \email{sayarkarmakar@ufl.edu}
\email{sayarkarmakar@ufl.edu}
}

\author{Sanchari Das}
\affiliation{%
  \institution{Indiana University \& University of Denver}
  }
\email{sancdas@iu.edu}
\newcommand{\ignore}[1]{}
\renewcommand{\shortauthors}{Karmakar and Das}

\begin{abstract}
COVID-19's impact has surpassed from personal and global health to our social life. In terms of digital presence, it is speculated that during pandemic, there has been a significant rise in cyberbullying. In this paper, we have examined the hypothesis of whether cyberbullying and reporting of such incidents have increased in recent times. To evaluate the speculations, we collected cyberbullying related public tweets ($N=454,046$) posted between January 1$^{st}$, 2020 -- June 7$^{th}$, 2020. A simple visual frequentist analysis ignores serial correlation and does not depict changepoints as such. To address correlation and a relatively small number of time points, Bayesian estimation of the trends is proposed for the collected data via an autoregressive Poisson model. We show that this new Bayesian method detailed in this paper can clearly show the upward trend on cyberbullying-related tweets since mid-March 2020. However, this evidence itself does not signify a rise in cyberbullying but shows a correlation of the crisis with the discussion of such incidents by individuals. Our work emphasizes a critical issue of cyberbullying and how a global crisis impacts social media abuse and provides a trend analysis model that can be utilized for social media data analysis in general.
\end{abstract}


\maketitle

\section{Introduction}
Bullying is characterized as the \lq\lq repeated oppression, psychological or physical, of a less powerful person by a more powerful one\rq\rq~~\cite{farrington1993understanding}. With the ascent of online communication, the dynamics of bullying have transcended beyond physical boundaries to the digital realm, referred to as \lq\lq cyberbullying\rq\rq~. Cyberbullying has gotten increasingly pervasive, as focused exploitation has moved from face to face to advanced stages, targeting users despite geographic constraints~\cite{ybarra2007examining, slonje2008cyberbullying}. Victims of cyberbullying can be targeted through various sources, including mobile phones, video cameras, emails, and web pages~\cite{ybarra2004online}. Cyberbullying can negatively impact mental health, with 32\% of victims reporting symptoms of stress and 38\% of victims experiencing emotional distress, even after the online abuse stops~\cite{finkelhor2000online, ybarra2006examining}. 

Earlier investigations have indicated that web-based social networking has expanded the impact of cyberbullying~\cite{ybarra2006examining}. On social networking sites and applications, cyberbullying is particularly common, with 66\% of all cyberbullying episodes occurring on these platforms~\footnote{\url{https://www.pewresearch.org/internet/2014/10/22/part-2-the-online-environment/}}. Twitter permits individuals to now and again interact with outsiders (counting celebrities)~\cite{das2017celebrities}; however, this also leads others to mirror and forge identities online and trick users~\cite{tsoutsanis2012tackling}. Verification of profiles only works for celebrities or those who are well-known in their field, making it difficult to verify an individual's identity~\cite{meligy2017identity}. It is even more challenging to identify abusers when they are imitating someone else. Due to the correlation of cyberbullying with social media usage, individuals often have shown negative user experience in these social media platforms~\cite{noman2019techies}.

Besides, with the current COVID-19 pandemic, individuals have increased their social media use to remain associated with others while social distancing~\cite{wiederhold2020social}. In any case, there have likewise been reports of incivility through such platforms~\cite{kim2020effects}. An abrupt ascent in internet-based life use - joined by children and adolescents continually utilizing such stages - could make a concerning spike in cyberbullying~\footnote{\url{https://www.digitaltrends.com/news/coronavirus-cyberbullying-separation learning/}}. Along these lines, our goal was to explore explicitly: \textit{How has a crisis, such as a pandemic (COVID-19), impacted reporting and discussions of cyberbullying incidents on Twitter?}

To understand users' perspectives, we collected $454,046$ of publicly available tweets about cyberbullying to understand user discussions online. We first tried a simple visual analysis to detect a significant rise in the incidence count of these keywords anytime around March. However, as one can see from Figure~\ref{fig:tc} or~\ref{fig:sub-class} such a changepoint is not very prominent. This allowed us to address the shortcomings of such a simplistic model, which ignores the possibility of a smooth change, the inherent dependence in the time series of counts. The initial analysis motivated our research to build a suitable autoregressive bayesian model, as described in Section~\ref{sec:bayesianmodel}. We choose a time-varying Bayesian method previously detailed by Karmakar et al.~\cite{sk2020}. As hypothesized, we noticed an increase in cyberbullying incident discussions during the pandemic, which shows an impact of the crisis on cyberbullying trends. Our method allowed us to construct posterior samples of the parameter function of time with the collected data. To the best of our knowledge, this work is the first quantitative trend analysis on a large sample data. Our results also reveal a clear telling effect of COVID-19 on worsening cyberbullying incidents as reported and discussed through tweets. Based on our quantitative work, we can explore in-depth qualitative analysis as a future extension of this work to further see the details of the discussed tweets.

\section{Related work}
\label{sec:rw}
Cyberbullying has expanded significantly with the advent of social media and billions of users being online everyday~\cite{slonje2008cyberbullying}. User experience of cyberbullying has been reported in several social networking platforms, chat rooms, and mobile messaging applications; such abuse transcends beyond geographical proximity~\cite{smith2006investigation,vogel2014social}. Furthermore, in light of the fact of crisis, it is being speculated that the crisis has increased cyberbullying incidents~\cite{depoux2020pandemic}. To address the issue of cyberbullying, prior research has investigated online maltreatment and created effective technical and policy-focused~\cite{das2018modularity,das2019privacy} mitigation strategies. However, though some of these strategies are being implemented through social media policy management, there are several privacy concerns of the social media users~\cite{ayaburi2020effect,dev2018privacy}. Thus, the speculation about the rise of cyberbullying due to a pandemic is a natural progression, which requires detailed analysis to verify such hypothesis. To understand further, we start by analyzing the cyberbullying discussion trends over twitter.

\subsection{Emergence and Effects of Cyberbullying}
Mason defined cyberbullying as \lq\lq an individual or a group willfully using information and communication involving electronic technologies to facilitate deliberate and repeated harassment or threat to another individual or group by sending or posting cruel text and/or graphics using technological means\rq\rq~ ~\cite{mason2008cyberbullying}. To investigate further, Nocentini et al. studied the behaviour of the attackers for different types of cyberbullying, including imbalance of power, intention, repetition, anonymity, and publicity~\cite{nocentini2010cyberbullying}. Previous works have explored the effects of cyberbullying on targets, especially on teenagers; sometimes such abuses can impact both the cyberaggressors and cybervictims~\cite{vsleglova2011cyberbullying,mchugh2017most,wisniewski2016dear,bonanno2013cyber}. Dredge et al. noted the detrimental effects of cyberbullying on the social and emotional lives of targets, with the severity of the impact of the harassment depending on different factors, including the anonymity of the perpetrators and the presence of bystanders~\cite{dredge2014cyberbullying}. All of the above-mentioned studies and several other researchers~\cite{patchin2010cyberbullying,kowalski2012cyberbullying} indicate the severity of cyberbullying on individuals and for the society, thus it is critical to develop strong defense against cyberbullying.

\subsection{Defence Against Cyberbullying}
\subsubsection{Technical Mitigation Techniques}
Several technical mitigation techniques have been proposed, with the goal of automatically detecting and intervening in cyberbullying incidents online~\cite{ashktorab2016designing, ashktorab2016study}. For instance, Dinakar et al. proposed an online dashboard, which would allow moderators to track potential bullying incidents on a forum through natural language processing~\cite{dinakar2012common}. Mondal et al. analyzed hate speech on Twitter and Whisper to improve automated detection of bullying~\cite{mondal2017measurement}, while Cortis and Handschuh analyzed bullying tweets in the context of major world events~\cite{cortis2015analysis}. Such technical mitigation strategies are helpful, yet it is also critical to understand an individual's perspective and any policy implementation strategies adapted by the social media organizations.

\subsubsection{Organizational Policies}
As a mitigating measure, some prior work has focused on improving social media policies to prevent perpetrators from abusing their victims~\cite{crawford2016flag}. Pater et al. compared the social media policies of 15 different platforms and found that these policies vary from mild censoring to the involvement of law enforcement~\cite{pater2016characterizations}. Given the legal implications, there can be severe consequences for such incidents for these social media organizations~\cite{milosevic2018protecting}. Milosevic examined the responsibilities of social media companies in addressing cyberbullying among children~\cite{milosevic2016social}. 

\subsubsection{User Perspective}
In order to improve anti-harassment measures, previous research has examined the motivations of cyberbullies~\cite{lai2016cyberbullying,difranzo2018upstanding,mcnally2018co,singh2017they}. Lee and Kim interviewed $110$ subjects to investigate why social media users leave benevolent or malicious comments~\cite{lee2015people}. Whittaker and Kowalski found that cyber aggression was more present in online comment sections and forum replies than on Facebook, again suggesting the importance of anonymity~\cite{whittaker2015cyberbullying}. 

Overall, these defensive mechanisms are helpful and aid in making internet and online experience better for individuals. But, even with such strong defensive tactics, it is speculated that cyberbullying has increased, especially during the COVID-19. Thus, for our work we try to understand the problem better through detailed quantitative analysis.

\subsection{Cyberbullying Trend Analysis}
Studies that analyze trends in cyberbullying are helpful in understanding how events can impact digital users. Schneider et al. conducted four surveys across 17 high schools and found that the overall rate of cyberbullying increased from 2006 to 2012~\cite{kessel2015trends}. Snell and Englander through survey-based analysis found that females are more likely to be involved in cyberbullying as both victims and as perpetrators, indicating the importance of gender as a factor in mitigating online bullying~\cite{snell2010cyberbullying}. Mangaonkar et al. used a distributed design for analyzing tweets and detecting cyberbullying in real-time ~\cite{mangaonkar2015collaborative}.

Twitter allows users to express themselves in 280 character \lq tweets;\rq~ prior studies have analyzed these messages for cyberbullying~\cite{alim2015analysis,nurrahmi2018indonesian}. Cortis and Handschuh analyzed bullying tweets in the context of two trending events (the Ebola outbreak and shooting of Michael Brown in Ferguson, Missouri) and identified commonly used hashtags and named entitites in bullying tweets~\cite{cortis2015analysis}. They tried to identify cyberbullies through these discussions, but whether or not such crisis situations increased bullying tweets was not studied. Due to an increase in individuals' online digital presence, assumptions have been made that the pandemic situation from COVID-19 can increase cyberbullying attacks. Thus, our goal was to find concrete evidence to support or contradict this hypothesis, also rather than surveys, we chose to collect data from Twitter itself to get the real-time trend of such critical incidents.

\section{Method: Data collection}
\label{sec:data}
With over 300 million active daily users, Twitter~\footnote{\url{https://twitter.com/login}} is an ideal data source~\footnote{\url{https://www.statista.com/statistics/282087/number-of-monthly-active-twitter-users/}}. Thus, to assess the impact of COVID-19 on cyberbullying, we collected $454,046$ public tweets on Twitter, all of which mentioned cyberbullying. We scraped Twitter for user-posted, publicly available tweets related to the topics of cyberbullying, social media bullying, online harassment, etc. The data was collected using Get Old Tweets API~\footnote{\url{https://github.com/Jefferson-Henrique/GetOldTweets-python}}, which allowed us to access tweets older than one week. This API was used in the web crawler written in Python, and the data was stored with MongoDB. The data collection spanned from January 1st 2020--June 7th 2020. This timeline was particularly selected to note the impact of COVID-19 on online users and determine whether the crisis situation led to an increase in online abuse. We used the following key terms when conducting our search: ~\textit{ Internet bullying, Internet bully, Internet bullies, online abuse, online harassment, online shaming, online stalking, cyberbullying, social media bullying, stop cyberbullying, cyberbully, cyberbullies, FB bullying, FB cyberbullying, FB harassment, FB victim, Facebook bullying, Facebook cyberbullying, Facebook victim, Facebook harassment, Twitter bullying, Twitter cyberbullying, Twitter harassment, Twitter victim, Insta bullying, Insta cyberbullying, Insta harassment, Insta victim}. We only collected direct tweets and removed any retweets or duplicate tweets. 

\section{Change-point analysis}
\begin{figure}[ht]
    \centering
    \includegraphics[width=80mm, height=45 mm]{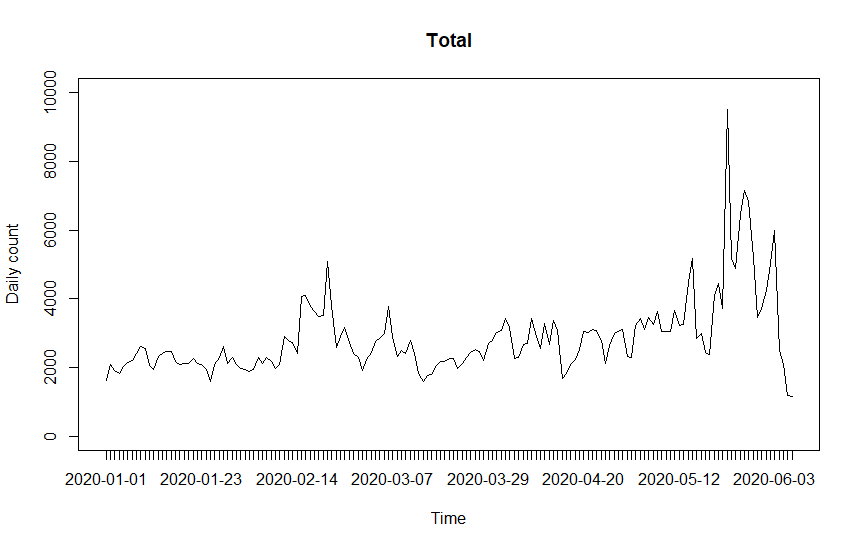}
          \caption{Daily count of total tweets related to bullying}
    \label{fig:tc}
    \vspace{-3mm}
\end{figure}

After completing the data collection, we performed trend analysis to evaluate the impact of the crisis situation, such as COVID-19 pandemic on cyberbullying. Using the timestamp of the post, we obtained the daily count of the tweets which including at least one of these keywords. Figure \ref{fig:tc} shows the daily count for the $159$ days from 01$^{st}$ January, 2020 to 07$^{th}$ June, 2020.

\begin{figure}[ht]
    \centering
    \includegraphics[width=80mm, height=65 mm]{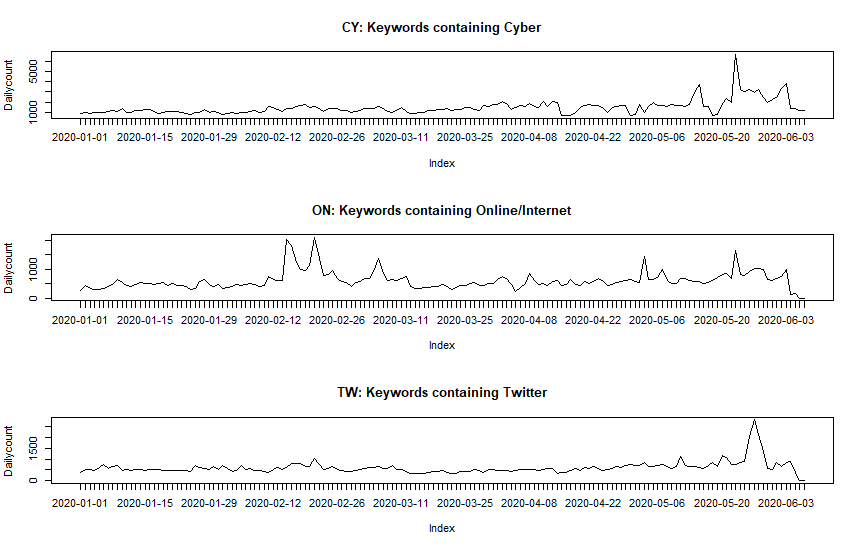}
          \caption{Daily count of total tweets for the three sub-class}
    \label{fig:sub-class}
    \vspace{-5mm}
\end{figure}

\noindent Some of the types of keywords had fewer tweets with negligible impact on the analysis. Thus, we broadly divide them in 3 sub-classes: keywords containing ``cyber'' (CY, 7 keywords, $235,542$ tweets), ``online/internet''(ON, 6 keywords, $96,629$ tweets) and ``twitter''(TW, 3 keywords $96,147$, tweets). The daily count distribution for these sub-classes are shown in Figure~\ref{fig:sub-class}. 

\subsection{Traditional Change-Point Analysis}

In a traditional change-point analysis, one looks for an abrupt change; i.e. after observing $X_1,\ldots, X_T$ if we suspect there is at most one change point then we are looking for the unknown location $1 \leq \tau \leq T$, such that 

\begin{eqnarray}\label{eq:cp}
E(X_i)= \mu\text{ if }i \leq \tau, \text{ and } E(X_i)= \mu+\delta\text{ if }i >\tau \text{ where } \delta \neq 0
\end{eqnarray}

The analysis shows a clear pattern that is prevalent to all the counts and the sub-classes that we present here. Overall, except for the sub-class \lq ON\rq~, there does not seem to be a huge change in mean except it went slightly upwards since mid-March and in all categories including the total. We notice a huge spike in the cyberbullying related tweets in the second half of May. The sudden rise in the frequency of tweets in the second half May, can be due to the untimely demise of the Japanese TV star~\footnote{\url{https://www.japantimes.co.jp/news/2020/05/30/national/media-national/online-harrassment/}} which occurred due to cyberbullying. Moreover for the class \lq ON\rq~, one can see a significant spike in the second half of February and the overall mean also had an upward trend. This may or may not be due to the pandemic. Note that except for the spike in later half of May, there is no abrupt break due to COVID-19. The authors explored  (\cite{daschange2020}) such a simple change-point model as an work in progress. However we believed that a simple model like (\ref{eq:cp}) can often fail to adequately capture some other sophistication that are particular to the data we collected here. 

\subsection{Short-coming of the simpler model}
\underline{Abrupt change vs Smooth change:} Note that the change-point model in \ref{eq:cp} address for abrupt change. However, due to the heterogeneous nature of Twitter, one can expect the change might not be abrupt and this can explain why just from the daily count summaries of the total tweets or the sub-classes do not reveal any abrupt change in either mean or variance in general. A more meaningful model could be where the parameters change smoothly over time and we can estimate these parameters as function of time and see whether the trend is increasing due to COVID-19 or not.  

\underline{Dependence:} Note that, the daily time count of number of occurrences is a time-series. Any visual analysis of change-point would heavily disregard the inherent dependence assumption that is present in a time-series. These counts depend heavily on current trend and are expected to show strong correlation with recent pasts. We decide to furnish this through the following Figure~\ref{fig:acf}. Under such heavy dependence for the total count and the three sub-series, one needs to take the dependence into account. Otherwise any analysis, be it abrupt change point or smooth time-varying parameter model will not be justified. 

\begin{figure}[ht]
    \centering
    \includegraphics[width=80mm, height=35 mm]{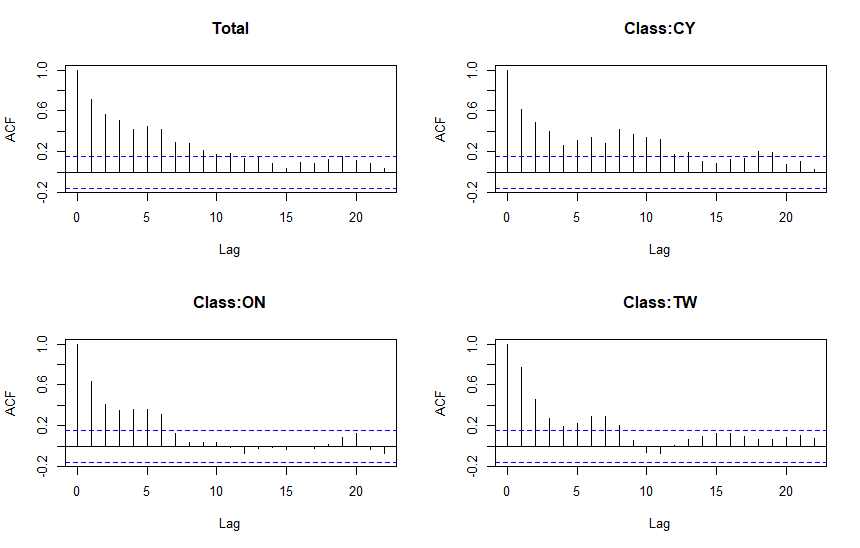}
          \caption{Daily count of total tweets for the three sub-class}
    \label{fig:acf}
    \vspace{-2mm}
\end{figure}

\underline{Poisson count time series:} Also note that the daily number of occurrences is a count series but unfortunately the traditional change-point analysis often assume normality (Normal distribution). Another advantage of using Poisson random variable is it can model the mean and variance through a single parameter. 

\underline{Small sample size:} A wide range of frequentist time-varying model were discussed in~\cite{sayar2020} and~\cite{karmakar2018asymptotic} that relied on kernel-based methods. However, one needs a large sample (Sample time point size of at least 500) to estimate any time-trend with precision. This is a sheer shortcoming of the kernel based methods that are essential for a frequentist model. But here we collected tweets of first 5 months of 2020 resulting in a sample of 159 time points.
\section{Smooth change: A Bayesian Model}
\label{sec:bayesianmodel}
In order to address the inadequacy of a visual detection of changepoint, we propose the following time-varying Bayesian autoregressive count (TVBARC) model from~\cite{sk2020}. Note that this model addresses the short-comings mentioned above.

\subsection{Model}
\label{ssc:model}
Due to the possibly non-stationary (over time) nature of the data, we propose a time-varying version of the linear Poisson autoregressive model \citep{zeger1988regression, brandt2001linear}. The conditional distribution for count-valued time-series $X_t$ given $\mathcal{F}_{t-1}=\{X_i: i\leq (t-1)\}$ is,
\begin{align}
    X_t|\mathcal{F}_{t-1}\sim& \mathrm{Poisson}(\lambda_t) \text{ where } \lambda_t=\mu(t/T) + \sum_{i=1}^p a_i(t/T) X_{t-i}.\label{TVBARC}
\end{align}
Due to the Poisson link in~\eqref{TVBARC}, both conditional mean and conditional variance depend on the past observations. The conditional expectation of $X_t$ in the above model \eqref{TVBARC} is $E(X_t|\mathcal{F}_{t-1})=\mu(t/T) + \sum_{i=1}^p a_i(t/T) X_{t-i}$, which is positive-valued. Additionally, we impose the following constraints on parameter space for the time-varying parameters, 
\begin{align}\label{eq:parcondition}
    \mathcal{P}_1=\{\mu, a_i:\mu(x)> 0, 0\leq a_{i}(x)\leq 1, \sup_{x}\sum_{k}a_{k}(x)<1\}.
\end{align}
When $p=0$, our proposed model reduces to routinely used nonparametric independent Poisson regression model as in \cite{shen2015adaptive}. The $\mu(\cdot)$ function correspond to the general mean trend at time $t$ and $a_p(\cdot)$, the $p$-th order autoregressive (AR hereafter) coefficient function denotes how the observation at time $t$ is affected by a past observation at lag $p$. The strong correlation pattern in Figure \ref{fig:acf} shows we should opt for a $p>0$.

\subsection{Posterior Computation}
\label{ssc:posterior}
To proceed with Bayesian computation, we put priors on the unknown functions $\mu(\cdot)$ and $a_i(\cdot)$'s such that they are supported in $\mathcal{P}_1$. The prior distributions on these functions are induced through basis expansions in B-splines with suitable constraints on the coefficients to impose the shape constraints as in $\mathcal{P}$. Detailed description of the priors are given below,
\begin{align*}
\mu(x) =&\sum_{j=1}^{K_1}\exp(\beta_j)B_j(x),  a_{i}(x)=\sum_{j=1}^{K_2}\theta_{ij}M_{i}B_j(x), \quad 0\leq\theta_{ij}\leq 1,\\
    M_i=&\frac{\exp(\delta_i)}{\sum_{k=0}^p{\exp(\delta_k)}}, \quad i=1,\ldots,p,\\
    \delta_l\sim&N(0, c_1),\textrm{ for }0\leq l\leq p,\beta_{j}\sim N(0, c_2)\textrm{ for } 1\leq j\leq K_1,\\
    \theta_{ij}\sim& U(0,1)\textrm{ for }1\leq i\leq p, 1\leq j\leq K_2\label{prior2}.
\end{align*}
Here $B_{j}$'s are the B-spline basis functions and $\delta_{j}$'s are unbounded. The prior induced by above construction are $\mathcal{P}$-supported. The verification is straightforward and can be found in \cite{sk2020}. We develop efficient MCMC algorithm to sample the parameter $\beta,\theta$ and $\delta$ from the above likelihood. Interested readers can see \cite{sk2020} for the computation of the likelihood and partial derivatives.

\section{Analysis of daily count data}
\label{sec:analysis}
\begin{figure}[ht]
    \centering
    \includegraphics[width=80mm, height=35 mm]{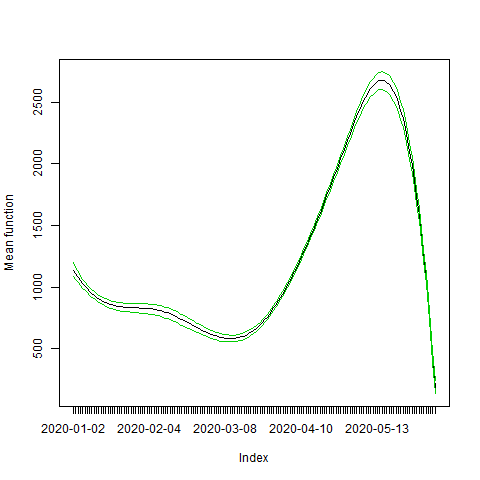}
          \caption{Total count: Mean trend of AR(1) model}
    \label{fig:total1mu}
    \vspace{-5mm}
\end{figure}

\begin{figure}[ht]
    \centering
    \includegraphics[width=80mm, height=35 mm]{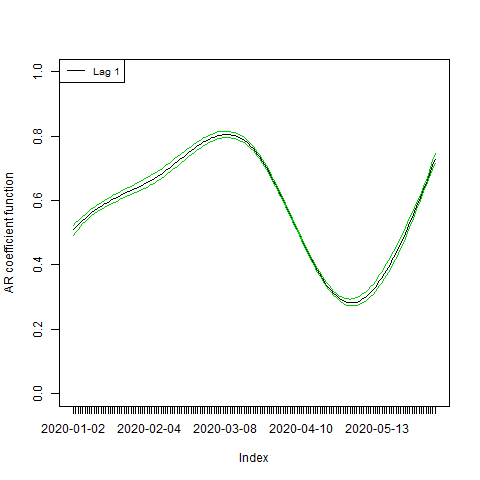}
          \caption{Total count: AR(1) trend of AR(1) model}
    \label{fig:total1A1}
   \vspace{-5mm}
\end{figure}

\begin{figure}[ht]
    \centering
    \includegraphics[width=80mm, height=35 mm]{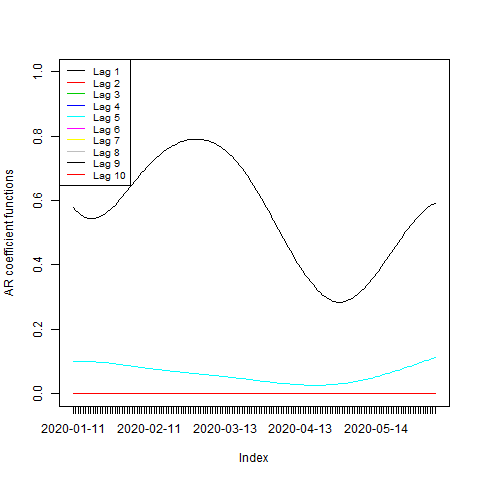}
          \caption{Total count: AR trends of AR(10) model}
    \label{fig:total10A1toA10}
   \vspace{-4mm}
\end{figure}

We first analyze the total count trends through two different choices of lag $p$: an AR(1) and an AR(10) model. Often there could be weekly patterns which could mean high correlation at lag 7 and also lag 8 if lag 1 was significant. To see if there is really a weekly pattern we decided to take a lag that is slightly higher than 7. The trend functions with their corresponding credible intervals (we omit the credible intervals for 10 AR coefficients for clarity) are shown in Figure \ref{fig:total1mu}, \ref{fig:total1A1} and  \ref{fig:total10A1toA10}. The trend and the credible intervals are mean and quantiles of 20000 posterior MCMC samples after 10000 burn-in.  

\noindent When we increase the number of lags to 10, we no longer report the credible intervals. From the above figures we summarize the findings as follows:
\begin{itemize}
    \item Mean trend increased from March 9th or so. If we contrast this with Figure \ref{fig:tc}, it is easy to appreciate the significant role of dependence for such time-series data.  
    \item For the AR(1) model, it increased upto March and then slowly decreased, it was again reaching a peak around end of May.  
    \item The AR(1) and AR(10) models are comparable and usually only the first lag accounts for most of the correlation. The mean trend $\mu(t)$ came out to be very similar to Figure \ref{fig:total1mu}. 
    \item The 95\% credible intervals provided are very narrow and thus gives us a significant confidence about the true trends being of similar nature.
\end{itemize}

\begin{figure}[!ht]
    \centering
    \includegraphics[width=80mm, height=100mm]{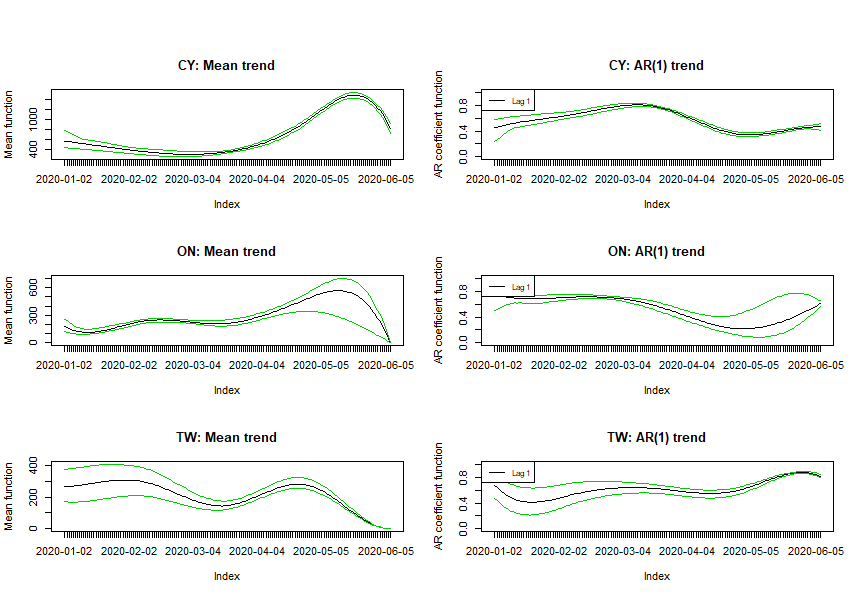}
          \caption{Sub-classes: Mean and AR(1) trends of model}
    \label{fig:tog}
    \vspace{-5mm}
\end{figure}

Next we analyze the three sub-classes under the AR(1) model and put them together in Figure \ref{fig:tog}. These sub-classes are very similar to the original total count. In all three of them there is a rise around first week of March and the rise continues for a while (except class TW). Nothing significant can be said about the AR coefficients except that the are also non-stationary. Note that with smaller number of data points for the classes ON and TW one can see the credible intervals are wider which is understandable. 

\section{Conclusion}
\label{sec:discussion}
Cyberbullying is a primary concern, and there have been several speculations on the increase in cyberbullying incidents during COVID-19. To start with this investigation, we perform a comprehensive Bayesian analysis of the daily count of cyberbullying occurrences and its dominant classes on cyberbullying-related public tweets ($N=454,046$) posted between January 1$^{st}$, 2020 -- June 7$^{th}$, 2020. We developed a Bayesian model that exhibited a sharp increase in the general mean trend for most of these twitter keywords related to cyberbullying. The significant AR correlation was at lag 1, and it evolved around 0.4-0.6 but not in a monotonic way. This analysis showed the increase in the cyberbullying discussion trend by Twitter users during COVID-19, which may or may not be due to the pandemic's direct impact. However, to further analyze the content discussed in these tweets, we plan to perform in-depth qualitative analysis. Our work is novel due to its first quantitative trend analysis on understanding the users based on their discussions regarding cyberbullying, especially during a pandemic crisis. Since COVID-19 has spread in multiple phases, it will be of utmost importance to detect such a high rise in social media trends. 

\begin{acks}
We would like to thank Umang Mehta for his help with the data collection. We would also like to acknowledge the support of Secure and Privacy Research in New-Age Technology (SPRINT) Lab, University of Denver; and Human and Technical Security (HATS) Lab, Indiana University, and University of Florida. Any opinions, findings, and conclusions or recommendations expressed in this material are solely those of the author(s). 
\end{acks}

\bibliographystyle{ACM-Reference-Format}
\bibliography{Bullyeicc}

\end{document}